\documentclass{aa}  
\usepackage{color}
\usepackage{graphicx}
\usepackage{lscape}
\usepackage{natbib}
\usepackage[varg]{txfonts}
\usepackage{url}
\usepackage{xspace}
\bibpunct{(}{)}{;}{a}{}{,}

\newcommand{\Teff}{$T\mathrm{\hspace*{-0.4ex}_{eff}}$\,}
\newcommand{\logg}{$\log\,g$\hspace*{0.5ex}}

\def\elf{\object{PG\,1159$-$035}}

\def\rx{\object{RX\,J0439.8$-$6809}}
\def\hh{\object{H1504+65}}
\begin{document}

\title{Analysis of HST/COS spectra of the bare C-O stellar core
  \hh\ and a high-velocity twin in the Galactic
  halo\thanks{Based on observations with the NASA/ESA Hubble Space
    Telescope, obtained at the Space Telescope Science Institute,
    which is operated by the Association of Universities for Research
    in Astronomy, Inc., under NASA contract NAS5-26666. }  }

\author{K\@. Werner
   \and T\@. Rauch
}

\institute{Institute for Astronomy and Astrophysics, Kepler Center for
  Astro and Particle Physics,  Eberhard Karls Universit\"at
  T\"ubingen, Sand~1, 72076 T\"ubingen,
  Germany\\ \email{werner@astro.uni-tuebingen.de} }

\date{Received 27 August 2015 / Accepted 25 September 2015}

\authorrunning{K. Werner \& T. Rauch} \titlerunning{The C-O
  atmospheres of the hot white dwarfs \hh\ and \rx}

\abstract{\hh\ is an extremely hot white dwarf (effective temperature
  \Teff = 200\,000\,K) with a carbon-oxygen dominated atmosphere
  devoid of hydrogen and helium. This atmospheric composition was
  hitherto unique among hot white dwarfs (WDs), and it could be
  related to recently detected cooler WDs with C or O dominated
  spectra. The origin of the H and He deficiency in \hh\ is
  unclear. To further assess this problem, we performed ultraviolet
  spectroscopy with the Cosmic Origins Spectrograph (COS) aboard the
  Hubble Space Telescope (HST).  In accordance with previous
  far-ultraviolet spectroscopy performed with the Far Ultraviolet
  Spectroscopic Explorer, the most prominent lines stem from
  \ion{C}{IV}, \ion{O}{V--VI}, and \ion{Ne}{VI--VIII}. Archival
  HST/COS spectra are utilised to prove that, considering its exotic
  composition, the supersoft X-ray source \rx\ is an even hotter
  (\Teff = 250\,000\,K) twin of \hh. In contrast to earlier claims, we
  find that the star is not located in the Large Magellanic Cloud, but
  is a foreground object in the Galactic halo at a distance of
  9.2\,kpc, 5.6\,kpc below the Galactic plane, receding with $v_{\rm
    rad}=+220$\,km\,s$^{-1}$. }


\maketitle

\section{Introduction}
\label{intro}

\hh\ was discovered as a very bright soft X-ray source by
HEAO1\footnote{High Energy Astronomy Observatory 1}
\citep{1983ApJS...51....1N} and its blue optical counterpart was
detected by \cite{1986ApJ...309..230N}. It was argued that the source
is an extremely hot white dwarf (\Teff = 160\,000\ $\pm$ 30\,000\,K
and surface gravity $\log$\,($g$/cm\,s$^{-2}$) = 7.5 $\pm$ 0.5) that  
is devoid of hydrogen and helium as a result of the lack of respective lines
in the optical spectrum. Non-local thermodynamic equilibrium (non-LTE)
modelling of ultraviolet (UV) and optical spectra proved the high
temperature and showed that the atmosphere is mainly composed by
carbon and oxygen in equal amounts \citep{1991A&A...251..147W}. Over
the years, substantial efforts were made with subsequent spectroscopy
in the soft X-ray range (with EUVE\footnote{Extreme Ultraviolet
  Explorer} and Chandra observatories), in the UV
(HUT\footnote{Hopkins Ultraviolet Telescope}, FUSE\footnote{Far
  Ultraviolet Spectroscopic Explorer}, HST\footnote{Hubble Space
  Telescope}), and in the optical with large ground-based telescopes
\citep{1999A&A...347L...9W, 2004A&A...421.1169W, 2004A&A...427..685W,
  2007A&A...474..591W, 2010ApJ...719L..32W}. These efforts lead to tight
constraints on effective temperature and gravity (\Teff = 200\,000
$\pm$ 20\,000\,K, \logg = 8.0 $\pm$ 0.5) and the detection of high
amounts of neon (2--5\,\%, by mass) and magnesium (2\,\%) as well as
iron at solar abundance level. High-resolution UV spectroscopy
presented in this paper was performed with the aim to further
constrain the trace element abundances and to shed more light on the
origin of the exotic composition of \hh.

Spectroscopically, \hh\ was classified as a PG\,1159 star, denoting
hot (\Teff = 75\,000--200\,000\,K), hydrogen-deficient (pre-)white
dwarfs (WDs) that suffered a late helium-shell flash, causing He-C
dominated atmospheres with high admixtures of oxygen \citep[He =
  0.30--0.85, C = 0.13--0.60, O =
  0.02--0.20,][]{2006PASP..118..183W}. \hh\ is an extreme member of
this class because of its helium-deficiency (He < 0.01, C = 0.5, O =
0.5). We argued that the extreme abundances are perhaps the
consequence of additional, strong mass loss, eroding the star down to
its C-O core or, alternatively, that the star is the result of a
binary WD merger. It could also be that this most massive known
PG\,1159 star ($M = 0.84^{+0.13}_{-0.10}\,M_\odot$) emerged from a
relatively massive main-sequence star ($8-10\,M_\odot$), which evolved
into a carbon-burning, super-AGB star
\citep[e.g.][]{1997ApJ...485..765G,2015MNRAS.446.2599D,2015ApJ...810...34W}. 

As to its future evolution, we speculated that \hh,  should it have
retained traces of helium in its envelope, would evolve into a
pure-He atmosphere (DB) white dwarf and later into a DQ white dwarf, which exhibits traces of C due to dredge-up by a thin, convective
He-envelope \citep{1982A&A...116..147K}. More drastic examples of
possible cooled-down versions of \hh\ were found recently, namely, the
so-called hot DQ white dwarfs \citep[\Teff =
  18\,000--24\,000\,K,][]{2007Natur.450..522D,2011ApJ...733L..19D}
with carbon-dominated atmospheres.  In addition, helium-dominated WDs
with O-rich atmospheres were discovered  and they are interpreted as
almost naked O-Ne stellar cores \citep[\Teff$\approx$ 10\,000\,K,
][]{2010Sci...327..188G,2015MNRAS.446.4078K}. 

Although the number of known PG\,1159 stars has increased to about
50 in the recent years, mainly as a result of the Sloan Digital Sky Survey
\citep{2014A&A...564A..53W}, \hh\ remained a unique object. However, another soft X-ray source was suspected of being  a
similar object since some time ago. \rx\ was discovered by ROSAT and classified as a
persistent supersoft X-ray source most probably located in the Large
Magellanic Cloud \citep[LMC;][]{1994A&A...281L..61G} and the blue
optical counterpart was detected by \cite{1996A&A...307L..49V}. The
spectral energy distribution was found to be consistent with a WD with
\Teff $\approx$\,300\,000\,K. A UV spectrum taken by
\cite{1999A&A...351L..27V} with HST/STIS\footnote{Space Telescope
  Imaging Spectrograph} is rather featureless. No flux variability was
detected. It was argued that the source is either a supersoft X-ray
binary powered by stable nuclear shell burning or a single hot
pre-white dwarf approaching the WD cooling sequence. The idea that the
source might be similar to \hh\ was corroborated with optical
spectroscopy \citep{2002ASPC..261..653R}, which revealed, among others,
\ion{O}{vi} emission lines. They were found redshifted by
$250\pm30$\,km\,s$^{-1}$.

In the following, we describe the HST observations
(Sect.\,\ref{sect:obs}) and model atmospheres (Sect.\,\ref{sect:models})
utilised for the spectral analysis, which  is presented in
Sects.\,\ref{sect:results} and \ref{sect:resultsrx} for \hh\ and \rx,
respectively. Stellar parameters and distances are derived in
Sect.\,\ref{sect:distances}. We summarise the results and conclude in
Sect.\,\ref{sect:conclusions}.

\begin{table}
\begin{center}
\caption{Observation log of HST/COS spectroscopy of \hh\ and \rx.\tablefootmark{a} }
\label{tab:obs} 
\tiny
\begin{tabular}{cccccc}
\hline 
\hline 
\noalign{\smallskip}
Star & Dataset & Grating & $\lambda$/\AA\ [A] & $\lambda$/\AA\ [B] & t$_{\rm exp}$/s  \\
\hline 
\noalign{\smallskip}
H1504   & LB3351010 & G130M & 1133--1279 & 1288--1433 & 1560 \\
        & LB3351020 & G160M & 1407--1587 & 1598--1777 & 2773 \\
J0439   & LCB501010 & G130M & 1133--1279 & 1288--1433 & 14080 \\ 
\noalign{\smallskip} \hline
\end{tabular} 
\tablefoot{\tablefoottext{a}{Columns 4 and 5 give the wavelength
    ranges covered by detector segments A and B. Exposure times are
    given in the last column.}} 
\end{center}
\end{table}

\section{Observations}
\label{sect:obs}

HST far-ultraviolet spectroscopy of \hh\ with the Cosmic Origins
Spectrograph (COS), using gratings G130M and G160M at central
wavelength settings 1291 and 1600\,\AA, respectively, was performed
during Cycle~17. They were conducted during two consecutive orbits on
May 25, 2010. Table\,\ref{tab:obs} summarises the observations. The
resolving power $\lambda/\Delta\lambda$ of the G130M and G160M
gratings across the observed wavelength ranges is 15\,000--19\,000 and
15\,000--21\,000, respectively, corresponding to $\Delta\lambda
\approx 0.08$ and 0.09\,\AA, respectively. The spectra were corrected
for the radial velocity of the star ($+44\pm4$\,km\,s$^{-1}$) for 
comparison with our synthetic spectra. As measured from the
Lyman~$\alpha$ wings, the hydrogen column density towards \hh\ is
N$_{\rm H\,I} = 6.0 \times 10^{19}$\,cm$^{-2}$, identical with a
result from previous HUT observations \citep{1998ApJ...502..858K}. The
spectrum is shown in Fig.\,\ref{fig:h1504}.

The HST/COS spectrum of \rx\ was retrieved from the MAST archive
(Proposal ID 13289). The observation was performed on Jan. 10, 2014
with the G130M grating in the same set-up as our observation of \hh,
with an exposure time of 14\,080\,s, however, only for one central
wavelength setting (1291\,\AA). The spectrum was corrected for the
radial velocity of the star ($v_{\rm rad}= +220\pm10$\,km\,s$^{-1}$)
for comparison with synthetic spectra. This velocity is consistent
with that measured from optical spectra ($+250\pm30$\,km\,s$^{-1}$;
see Introduction) and is close to the radial velocity of the LMC
\citep[+262 km\,s$^{-1}$;][]{2012AJ....144....4M}. The low-ionisation
interstellar absorption lines (\ion{C}{ii}, \ion{Si}{ii}, \ion{S}{ii},
\ion{Fe}{ii}) have $v_{\rm rad}=+8\pm 5$\,km\,s$^{-1}$. The strongest
of them display a second, much weaker velocity component with $v_{\rm
  rad}= +150 \pm 10$\,km\,s$^{-1}$, which probably stems from a
high-velocity cloud in the Galactic halo \citep[see
  e.g.][]{1999Natur.402..386R,2015MNRAS.451.4346S}. These
high-velocity interstellar components are not uncommon in UV spectra
of very hot pre-white dwarfs, as, for instance, encountered in a recent
study of four H-deficient objects with distances 2--8\,kpc
\citep{2014A&A...566A.116R}. One of these (HS\,1522+6615) is a halo
object 7.9\,kpc away (5.5\,kpc from the Galactic plane) with a high
radial velocity of $-180$\,km\,s$^{-1}$. 

The spectrum  of \rx\ is contaminated by strong airglow lines. Besides
Lyman~$\alpha$, there are strong emissions from \ion{N}{i} (at
1134/1135 and 1200\,\AA) and \ion{O}{i} (at 1302/1306 and
1356/1358\,\AA); see the airglow line table on the COS instrument
page\footnote{\url{http://www.stsci.edu/hst/cos/calibration/airglow_table.html}}. As
measured from the Lyman~$\alpha$ wings, the interstellar hydrogen
column density towards \rx\ is N$_{\rm H\,I} = 3.5 \times
10^{20}$\,cm$^{-2}$, consistent with the previous result from a
HST/STIS observation \citep[$4 \times
  10^{20}$\,cm$^{-2}$,][]{1999A&A...351L..27V}. The spectrum is
shown in Fig.\,\ref{fig:rxj0439}.

\begin{table}
\begin{center}
\caption{Number of levels and lines of model ions used for line
  formation calculations of metals.\tablefootmark{a} }
\label{tab:modelatoms} 
\tiny
\begin{tabular}{cccccccccc}
\hline 
\hline 
\noalign{\smallskip}
   & III &  IV    &   V   &    VI  &   VII   & VIII   \\ 
\hline 
\noalign{\smallskip}
C  & 1,0 & 54,291 &   \\   
N  &     & 16,30  & 54,297\\   
O  &     & 1,0    & 12,18 & 54,291 \\
Ne &     &        &       & 92,687 & 103,761 & 77,510 \\
Mg &     &        & 15,18 & 27,60  & 46,147  & 50,269 \\
Si &     &        & 25,59 & 45,193 & 61,138  & 55,239 \\
\hline 
\noalign{\smallskip}
   & VIII  & IX & X & XI \\ 
\hline 
\noalign{\smallskip}
Ca & 1,0   & 15,23 & 25,126 & 4,2 \\
\noalign{\smallskip} \hline
\end{tabular} 
\tablefoot{ \tablefoottext{a}{First and second number of each table
    entry denote the number of levels and lines, respectively. Not
    listed for each element is the highest ionisation stage, which
    only comprises its ground state. For the treatment of iron, see
    text.  }  } 
\end{center}
\end{table}

\begin{landscape}
\addtolength{\textwidth}{6.3cm}  \addtolength{\evensidemargin}{0cm}
\addtolength{\oddsidemargin}{0cm}
\begin{figure*}[bth]
  \centering
  \includegraphics[width=0.67\textwidth,angle=-90]{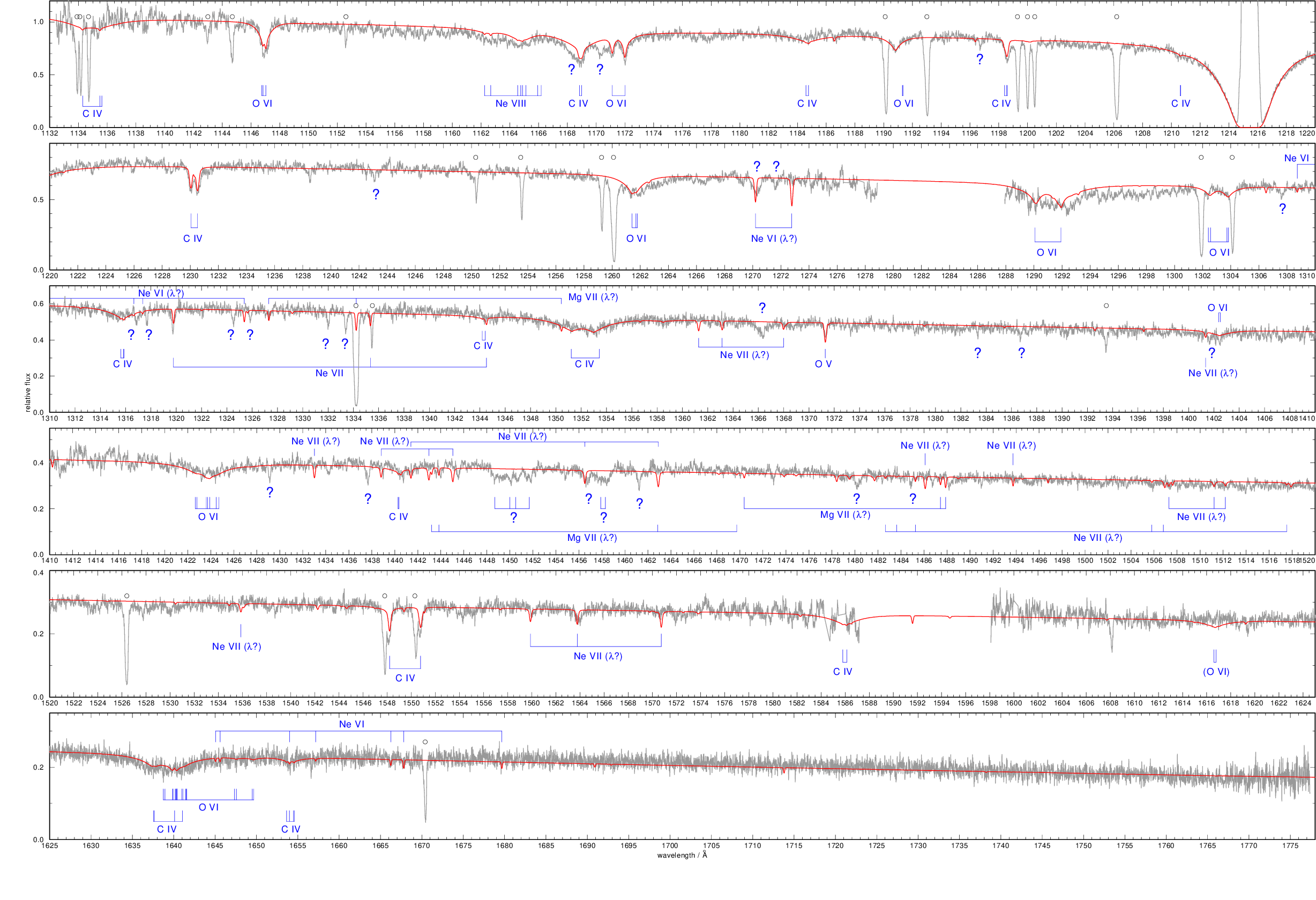}
  \caption{HST/COS spectrum of \hh\ and a model (\Teff = 200\,000\,K,
    \logg = 8,  C = 0.46, O = 0.46, Ne = 0.06, Mg = 0.02). Observation
    and model were folded with 0.02\,\AA\ and 0.1\,\AA\ Gaussians,
    respectively. Unidentified photospheric lines are marked with
    ''?''. Other observed narrow lines are of interstellar
    origin (the strongest are marked with circles). ''$\lambda$?'' 
    denotes lines in the model with uncertain
    wavelength position. }\label{fig:h1504}
\end{figure*}
\end{landscape}

\begin{landscape}
\addtolength{\textwidth}{6.3cm}  \addtolength{\evensidemargin}{0cm}
\addtolength{\oddsidemargin}{0cm}
\begin{figure*}[bth]
  \centering
  \includegraphics[width=0.67\textwidth,angle=-90]{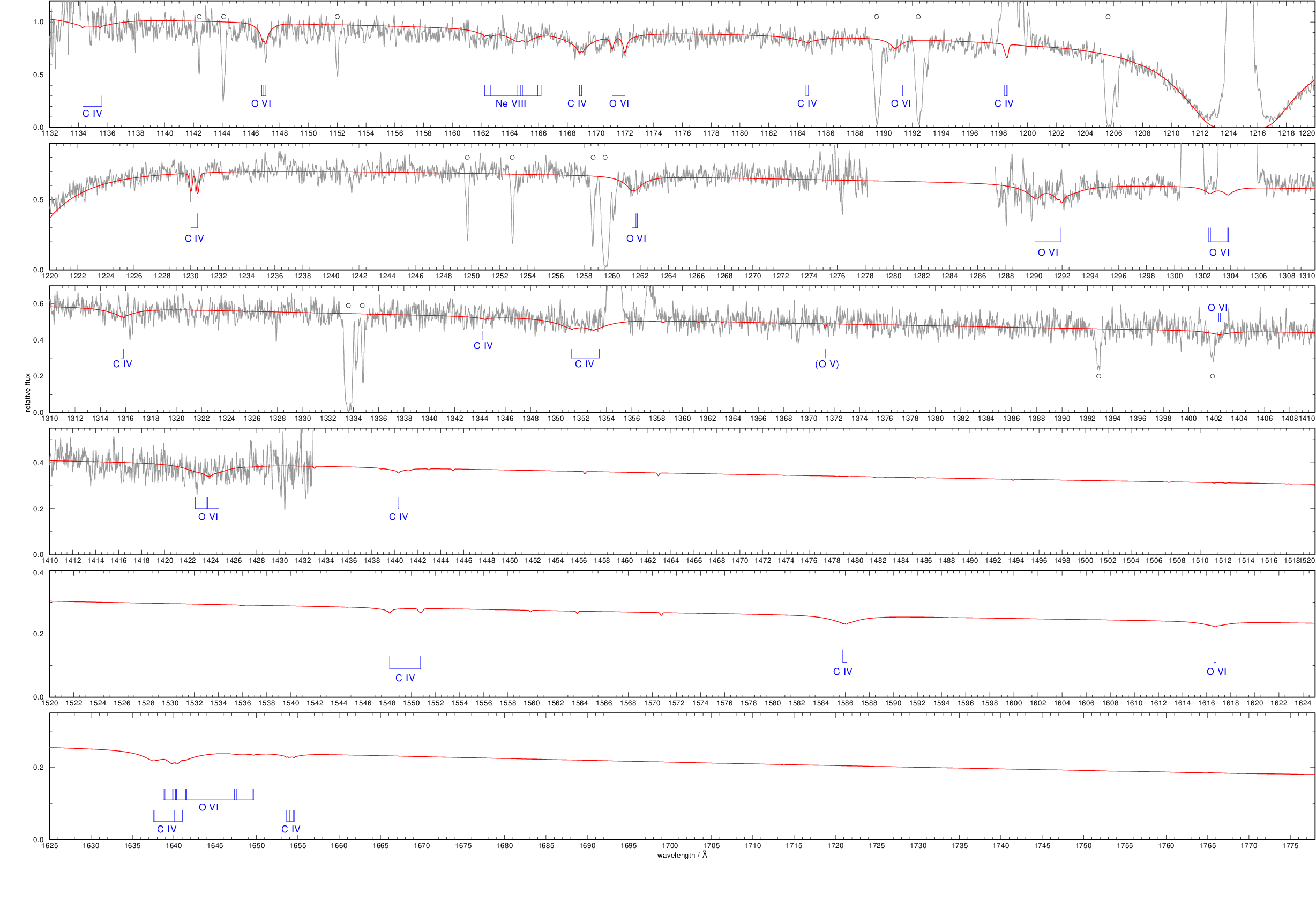}
  \caption{HST/COS spectrum of \rx\ and a model (\Teff = 250\,000\,K,
    \logg = 8,  C = 0.495, O = 0.495, Ne = 0.01). Observation and
    model were folded with 0.05\,\AA\ and 0.1\,\AA\ Gaussians,
    respectively. Unidentified photospheric lines are marked with
    ''?''. Other observed narrow lines are of interstellar origin 
   (the strongest are marked with circles).
  }\label{fig:rxj0439}
\end{figure*}
\end{landscape}

\begin{table}
\begin{center}
\caption{Identified photospheric lines in the
  HST/COS spectrum of \hh.\tablefootmark{a} }
\label{tab:identified-lines-cos} 
\tiny
\begin{tabular}{lcc}
\hline 
\hline 
\noalign{\smallskip}
Wavelength / \AA & Ion & Transition \\ \hline
\noalign{\smallskip} 
1134.25, 1134.30 & \ion{C}{iv}   & 4d -- 9f \\
1135.50          & \ion{C}{iv}   & 4f -- 9g \\
1146.75, 1146.83, 1477.03 & \ion{O}{vi}   & 4d -- 5p \\
1162.24, 1162.67 & \ion{Ne}{viii}& 5d -- 6f \\
1164.54, 1164.76 & \ion{Ne}{viii}& 5f -- 6g \\
1164.88          & \ion{Ne}{viii}& 5g -- 6h \\
1168.85, 1168.99 & \ion{C}{iv}   & 3d -- 4f \\
1171.12, 1172.00 & \ion{O}{vi}   & 4p -- 5s \\
1171.60:\tablefootmark{b} & \ion{Ne}{viii}& $\mathrm{6h\ ^2H^o}$ -- $\mathrm{8i\ ^2I}$ \\
1184.59, 1184.77 & \ion{C}{iv}   & 4p -- 8d  \\
1190.80, 1190.84 & \ion{O}{vi}   & 5s -- 7p  \\ 
1198.55, 1198.59 & \ion{C}{iv}   & 3d -- 4p  \\ 
1210.61:, 1210.65: & \ion{C}{iv}   & 4s -- 7p \\ 
1230.04, 1230.52 & \ion{C}{iv}   & 3p -- 4s  \\ 
1261.42, 1261.70, 1261.80 & \ion{O}{vi}   & 5p -- 7d \\
1270.3:\tablefootmark{c}  & \ion{Ne}{vi}  & $2{\rm s}2{\rm p}\ ^2{\rm F}^{\rm o} - 2{\rm s}^2\ ^2{\rm G}$\\ 
1290.06, 1290.18, 1290.21 & \ion{O}{vi}   & 5d -- 7f  \\
1291.81, 1291.84 & \ion{O}{vi}   & 5f -- 7g \\
1291.90, 1291.92 & \ion{O}{vi}   & 5g -- 7h \\
1292.00, 1292.02 & \ion{O}{vi}   & 5g -- 7f \\
1292.64, 1292.67, 1293.04 &\ion{O}{vi}   & 5f -- 7d \\
1298.\tablefootmark{g}                     & \ion{O}{vi}  & 6 -- 11 \\  
1302.41, 1302.46, 1302.56 & \ion{O}{vi}  & 5d -- 7p \\
1303.72, 1303.84 & \ion{O}{vi}   & 5p -- 7s \\
1315.62, 1315.85 & \ion{C}{iv}   & 4p -- 7d \\ 
1319.78 & \ion{Ne}{vii}& $\mathrm{2p\ ^1P^o}$ -- $\mathrm{2p^{2\ 3}P}$  \\
1344.18:, 1344.42:& \ion{C}{iv}   & 4p -- 7s \\  
1351.21, 1351.29 & \ion{C}{iv}   & 4d -- 7f \\ 
1352.97 & \ion{C}{iv}   & 4f -- 7g \\ 
1353.43:& \ion{C}{iv}   & 4f -- 7d \\  
1358.42:, 1358.47:, 1358.50: & \ion{C}{iv}   & 4d -- 7p \\ 
1371.29 & \ion{O}{v}    & $2{\rm s}\ ^1{\rm P}^{\rm o}$ -- $2{\rm p}\ ^1{\rm D}$ \\
1402.38:, 1402.47:, 1402.52: & \ion{O}{vi}   & 6p -- 10d\\
1422.65, 1422.79, 1422.81 & \ion{O}{vi}   & 6d -- 10f \\
1423.61, 1423.65 & \ion{O}{vi}   & 6g -- 10h \\
1423.90, 1423.92 & \ion{O}{vi}   & 6h -- 10i \\
1440.30, 1440.38 & \ion{C}{iv}   & 4s -- 6p \\ 
1548.20, 1550.77 & \ion{C}{iv}   & 2s -- 2p \\ 
1560.2:, 1563.9:, 1572.0:\tablefootmark{c} & \ion{Ne}{vii} & $\mathrm{5f\ ^3F^o}$ -- $\mathrm{6g\ ^3G}$ \\ 
1585.81:, 1586.11:, 1586.14:\tablefootmark{e} & \ion{C}{iv}   & 4p -- 6d \\ 
1637.54, 1637.65 & \ion{C}{iv}   & 4d -- 6f \\ 
1638.73, 1638.91, 1638.94 & \ion{O}{vi} & 6d -- 9f \\
1639.85, 1639.91 & \ion{O}{vi}&  6f -- 9g \\
1640.10 & \ion{C}{iv}   & 4f -- 6g \\ 
1640.32, 1640.36 & \ion{O}{vi} & 6h -- 9i \\
1640.97, 1641.09, 1641.10 & \ion{O}{vi} & 6f -- 9g \\
1645.06:, 1645.59, 1654.01:\tablefootmark{f}, \\
1657.16:, 1666.24, 1667.82, \\
1679.67  & \ion{Ne}{vi}  & $3{\rm s}\ ^4{\rm P}^{\rm o} - 3{\rm p}\ ^4{\rm P}$ \\
1653.63, 1653.99 & \ion{C}{iv}   & 4p -- 6s \\  
1654.46, 1654.57 & \ion{C}{iv}   & 4d -- 6p \\  
\noalign{\smallskip} \hline
\end{tabular} 
\tablefoot{
\tablefoottext{a}{Colons denote uncertain detections;}
\tablefoottext{b}{blend with \ion{O}{vi;}}
\tablefoottext{c}{see text;}
\tablefoottext{e}{at end of spectral segment;}
\tablefoottext{f}{blend with \ion{C}{iv;}}
\tablefoottext{g}{weak, broad depression; not included in model.}
} 
\end{center}
\end{table}

\section{Model atoms and model atmospheres}
\label{sect:models}

For the spectral analysis, we used our non-LTE code
TMAP\footnote{\url{http://astro.uni-tuebingen.de/~TMAP}}
\citep{tmap2012} to compute plane-parallel, line-blanketed atmosphere
models in radiative and hydrostatic equilibrium
\citep{1999JCoAM.109...65W,2003ASPC..288...31W}. These include the
three most abundant elements in \hh, namely, C, O, and Ne. Five more
species (N, Mg, Si, Ca, Fe) were investigated and treated one by one
as trace elements, i.e. keeping  the atmospheric structure fixed. In
the same manner, an extended model atom for Ne was introduced, meaning
that non-LTE population numbers were also computed for highly-excited
levels, which  were treated in LTE during the preceding model-atmosphere
computations. Table\,\ref{tab:modelatoms} summarises the number of
considered non-LTE levels and radiative transitions between them. All
model atoms were built from the publicly available T\"ubingen Model
Atom Database
(TMAD\footnote{\url{http://astro.uni-tuebingen.de/~TMAD}}), comprising
data from different sources, namely, \citet{1975aelg.book.....B}, the
databases of the National Institute of Standards and Technology
(NIST\footnote{\url{http://www.nist.gov/pml/data/asd.cfm}}), the
Opacity Project
\citep[OP\footnote{\url{http://cdsweb.u-strasbg.fr/topbase/topbase.html}},][]{1994MNRAS.266..805S},
the IRON project \citep{1993A&A...279..298H},
CHIANTI\footnote{\url{http://www.chiantidatabase.org}}
\citep{1997A&AS..125..149D,2013ApJ...763...86L}, and the
Kentucky Atomic Line
List\footnote{\url{http://www.pa.uky.edu/~peter/atomic}}. 

For iron, we used a statistical approach employing typically seven
superlevels per ion linked by superlines together with an opacity
sampling method \citep{1989ApJ...339..558A,2003ASPC..288..103R}.
Ionisation stages {\tiny VII--IX} augmented by a single, ground-level
stage {\tiny X} were considered per species. We used the complete line
list of Kurucz \citep[so-called LIN lists, comprising about
  $2.6\times10^5$ lines of the considered
  ions;][]{kurucz1991,kurucz2009, kurucz2011} for the computation of
the non-LTE population numbers, and the so-called POS lists (these
include only the subset of lines with precisely known, experimentally
observed line positions) for the final spectrum synthesis.

\section{Line identifications and spectral fitting for \hh}
\label{sect:results}

Generally, our new HST/COS spectrum is dominated by lines from
\ion{C}{iv} and \ion{O}{vi}, which  are well known from, e.g. the
PG\,1159 prototype \elf\ \citep{2007A&A...462..281J}. In addition, we
see lines from \ion{Ne}{vi-viii}. A complete list of identified lines
is listed in Table\,\ref{tab:identified-lines-cos}.

In the following, we present a detailed description of the spectrum and
our line fitting procedure. We began our analysis with a reference
model with literature values: \Teff = 200\,000\,K, \logg = 8, and a
C/O ratio of unity. It turned out that this model already matches  the observation very
well. In particular, the neon and oxygen ionisation
balances confirm the temperature within the previous error range ($\pm
20\,000$\,K). There is no necessity to revise the C/O ratio or the
surface gravity. 

\subsection{Carbon}

The depth of the \ion{C}{iv} resonance doublet is not quite achieved
by the 200\,000\,K model and would favour a slightly lower temperature
(180\,000\,K). But then the low-ionisation features of oxygen
(\ion{O}{v}~1371\,\AA) and neon (\ion{Ne}{vi} lines) become slightly too
deep, while the \ion{Ne}{viii} lines become too weak. The other
\ion{C}{iv} lines match well with the exception of the 1198 and
1230\,\AA\ lines, whose wings were significantly too broad. This fact
highlights a general problem with the line broadening theory. Usually,
the subordinate \ion{C}{iv} (and \ion{O}{vi}) line profiles are well
described by an approximate treatment for linear Stark broadening
because they are one-electron systems. This approximation is good for
lines involving levels with high principal and angular quantum numbers
($n$ and $l$) because the levels are nearly degenerate. For low-$l$
and low-$n$ levels this treatment becomes questionable. The 1198 and
1230\,\AA\ lines are such cases; they are the 3d--4p and 3p--4s
transitions, respectively. On the other hand, the 3d--4f
transition at 1169\,\AA,\ involving the high-$l$ 4f level, is matched
very well by the model. We note that the depression in its blue wing is
probably due to an unidentified line. We now treat the 1198 and
1230\,\AA\ lines with quadratic Stark effect and indeed, they fit the
observations much better.

\subsection{Nitrogen}

From the lack of the \ion{N}{v} 1239/1243\,\AA\ resonance doublet we
can infer an upper limit of N $<3.0 \times 10^{-5}$, which is much
tighter than  the previous limit derived from the absence of
optical \ion{N}{v} emission lines \citep[$<5.0 \times
  10^{-3}$,][]{1991A&A...251..147W}.

\subsection{Oxygen}

The detection of the \ion{O}{v}~1371\,\AA\ line is important because
it is very temperature sensitive. Its profile is best matched by a
\Teff = 200\,000\,K model, while a model with 180\,000\,K has a profile that is
definitely too deep and the line has vanished in a 220\,000\,K model .

We found that the Ritz wavelengths of the \ion{O}{vi} 5s--7p doublet
(1191.34 and 1191.30\,\AA) must be reduced by $0.5$\,\AA\ to match the
observed wavelengths, indicative of inaccuracies in the level energies
in the atomic databases utilised. Similar corrections for a number of
other UV \ion{O}{vi} lines were made by \citet{2007A&A...462..281J}
using the spectra of several PG\,1159 stars (see their Tab.\,6).

\subsection{Neon}

Neon lines of three ionisation stages (\ion{Ne}{vi-viii}) can be
identified. Our models predict a very strong \ion{Ne}{vi} line
at 1223.3\,\AA\ from the transition $5{\rm d}\ ^2{\rm D} - 7{\rm
  f}\ ^2{\rm F}^{\rm o}$, which is not present in the observation. The
line was computed with an f-value from OP and NIST level energies. The
upper level, however, is marked as ``may not be real'', so in fact the
transition might not exist at all. Consequently, we omitted it from
our synthetic spectra.

The next strongest \ion{Ne}{vi} lines in our models are the $2{\rm
  s}2{\rm p}3{\rm d}\ ^2{\rm F}^{\rm o} - 2{\rm s}^2 5{\rm g}\ ^2{\rm
  G}$ doublet at 1272\,\AA. According to \citet{1999JOSAB..16.1966K},
they are located at a single observed wavelength (1271.8$\pm$1.5\,\AA)
but two (computed) Ritz wavelengths (1270.2/1272.8$\pm0.5$\,\AA) and
both components have the same strength. The \hh\ spectrum has one
strong line at 1270.3\,\AA\ plus two nearby, weaker lines with similar
strength at 1269.3 and 1271.6\,\AA. The weaker lines might be assigned
to the modelled doublet because their spacing of 2.3\,\AA\ is similar
to that of the Ritz wavelengths (2.6\,\AA), although the absolute
differences (0.9 and 1.2\,\AA) are about twice the quoted uncertainty
of 0.5\,\AA. Another possibility is that the doublet spacing computed
by \citet{1999JOSAB..16.1966K} is overestimated and indeed
unresolvable in the \hh\ spectrum. The strong feature at
1270.3\,\AA\ coincides with the (unresolved) laboratory measurement of
\citet{1999JOSAB..16.1966K} at 1271.8\,\AA\  within its
1.5\,\AA\ uncertainty. This is also favoured by the fact that the total
equivalent width of the computed doublet better fits to the strong observed feature
than to the weaker observed lines. Another \ion{Ne}{vi}
multiplet at 1645--1680\,\AA\ is much weaker but the wavelengths of
the components are well known, \citep[$\pm$0.03\,\AA,
][]{1999JOSAB..16.1966K} and the strongest components are detectable
in the observation.

As to \ion{Ne}{vii}, our models predict many lines, however, because of
uncertain level energies their positions are not known sufficiently
well to be assigned to any of the unidentified observed lines. The
only line of this ion with accurately known wavelength is an
intercombination line at 1319.71\,\AA, and this line can be identified in the
observed spectrum. A triplet in our synthetic spectrum at
1560--1571\,\AA\ might be assigned to nearby absorption lines in the
observation with differences of about 1\,\AA.

The broad absorption trough at 1162--1166\,\AA\ stems from numerous
\ion{Ne}{viii} lines \citep{2007A&A...474..591W}.
The presence of lines from three ionisation stages of neon offers a
particularly strong constraint on \Teff. We confirm our earlier result
of \Teff = $200\,000\pm20\,000$\,K. The neon abundance is best matched
with a model with Ne = 0.06, which is  consistent with, though higher
than, earlier results.

\subsection{Silicon}

At \Teff = 200\,000\,K, silicon lines with accurately known
wavelengths from ionisation stages \ion{Si}{v-vi} are predicted by the
models. Among the strongest  are, for example, \ion{Si}{v}
1251.39 and \ion{Si}{vi} 1229.01\,\AA. None of these are  identified in the
observation. In addition, \ion{Si}{vii} lines are also predicted
(e.g. at 1236\,\AA), however, their wavelengths are uncertain by up to
1\,\AA. Nevertheless, no possible counterparts are detectable in the
observation. We derive an upper abundance limit of Si $< 2.6 \times
10^{-3}$, which is four times the solar value. An unidentified line at
1148.6\,\AA\ is well matched by a \ion{Si}{vi}  $^4{\rm P} - {^4{\rm D}}^{\rm o}$
line in the model with this abundance, however, this is perhaps
by a chance coincidence because the wavelength is known only within $\pm
0.12$\,\AA\ and another line of this multiplet at 1152.8\,\AA\ is not
unambiguously detected.

\subsection{Magnesium}

We searched for Mg lines without success. According to our models,
\ion{Mg}{vii} has the strongest lines, however, wavelength positions
are not known better than about 1--2\,\AA. At the abundance Mg = 0.02
derived from the Chandra soft-X-ray spectrum \citep{2004A&A...427..685W},
UV lines of significant strength (up to 15\% central depression) are
predicted, most prominently the components of a $\rm{^3P^o} -
\rm{^3P}$ triplet at 1291--1350\,\AA. Hence, some of the unidentified
lines could stem from this ion.

\subsection{Calcium}

The strongest line predicted by our models is \ion{Ca}{x} 1159.2\,\AA,
a component of the 4p--4d doublet.  We looked for this line
in the FUSE spectrum of \hh, consistent with an upper abundance limit
of solar \citep[Ca $<6.4\times10^{-5}$,][]{2008A&A...492L..43W}, to no avail. The
same holds for the COS spectrum where the line is not detectable either. We had tentatively detected the 4s--4p lines at 1461.2 and
1503.6\,\AA\ in the hot PG\,1159 central star of NGC\,246. In the COS
spectrum of \hh\ there are lines at these positions, but it is
unlikely that they stem from \ion{Ca}{x}. At solar Ca abundance, our
200\,000\,K model provides  lines that are too weak. Even a match of the weaker
component at 1503.6\,\AA\ would require Ca ten times solar, clearly at
odds with the non-detection of the 1159.2\,\AA\ line. Consequently, we
confirm the upper limit of solar.

\subsection{Iron}

 A solar iron abundance was derived
from \ion{Fe}{x} lines in FUSE data \citep{2010ApJ...719L..32W}. Our models do not predict detectable
lines in the COS spectral range. In particular, an unidentified
feature at 1619.4\,\AA\ is not a \ion{Fe}{viii} line in the Kurucz
list located at 1619.35\,\AA.
Estimated errors for the abundances are $\pm20\%$ for C and O and a
factor of 3 for the other elements \citep[see
  also][]{2004A&A...421.1169W}.

\begin{table}
\begin{center}
\caption{Unidentified photospheric lines in the HST/COS spectrum of \hh\ and possible identifications. }
\label{tab:unidentified-lines-cos} 
\tiny
\begin{tabular}{lll}
\hline 
\hline 
\noalign{\smallskip}
Wavelength / \AA & Possible id. & Remark\\ \hline
\noalign{\smallskip} 
1148.6 & \ion{Si}{vi}\\
1159.3 & \ion{Ca}{x} \\  
1168.5 & broad dip in \ion{C}{iv} wing \\
1170.3 & & strong \\
1196.7 \\            
1207.5 \\
1223.0 & \ion{Si}{vii} \\  
1225.2 \\  
1225.9 \\  
1228.3   & \ion{Si}{vii} \\  
1228.7   & \ion{Si}{vii} \\  
1243.1   & \ion{Si}{vii} \\  
1246.4 \\
1270.3 & \ion{Ne}{vi} & strong \\ 
1271.6 & \ion{Ne}{vi} &  \\ 
1307.6  &&  \tablefootmark{a} \\
1316.9   & \ion{Ne}{vi} \\  
1317.7   & \ion{Ne}{vi} \\  
1324.5 & & strong\\  
1325.8 & & strong \\ 
1332.0   & \ion{Na}{vii} & strong\\  
1333.4 && strong\\
1334.8 \\
1366.4   & \ion{Si}{vii} & strong\tablefootmark{b} \\
1383.4 \\  
1386.7 \\
1401.6   & \ion{Ne}{vii} \\  
1416.3   & \ion{Ne}{vii} \\  
1429.0   &               & strong\tablefootmark{c}  \\
1434.3   & \ion{Ne}{vii} \\  
1435.8  \\ 
1437.6   &               & strong \\
1439.0   & \ion{Ne}{vii} \\  
1443.6   & \ion{Ne}{vii} \\  
1444.0  \\ 
1445.5   & \ion{Ne}{vii} \\  
1449--1452 &              & strong broad dip, many lines \\
1454.8 \\
1456--1459 &              & strong broad dip, many lines \\
1461.2   &               & strong\tablefootmark{c} \\
1472.9   & \ion{Ne}{vii} \\  
1475.3   & \ion{Mg}{vii} \\  
1480.20  &               & strong\tablefootmark{b} \\
1484.9   & \ion{Ne}{vii} \\  
1486.75  & \ion{Ne}{vii} \\  
1488.2   & \ion{Mg}{vii}, \ion{Ne}{vii} \\ 
1491.0   & \ion{Ne}{vi}  & \tablefootmark{c} \\
1495.3   & \ion{Ne}{vii} \\  
1503.6   & \ion{Ca}{x} \\  
1538.7   & \ion{Ne}{vi} \\  
1543.6   & \ion{Ne}{vi} \\  
1554.5   &               & broad dip \\
1619.4   & \ion{Ne}{vii} \\  
\noalign{\smallskip} \hline
\end{tabular} 
\tablefoot{
\tablefoottext{a}{Also present in \elf\ and hot H-rich central stars;}
\tablefoottext{b}{also present in hot PG\,1159 stars;}
\tablefoottext{c}{also present in the hot PG\,1159 central star of NGC\,246.}
} 
\end{center}
\end{table}

\subsection{Unidentified lines}

A number of photospheric lines remains unidentified. The strongest of
them are listed in Table \,\ref{tab:unidentified-lines-cos} along
with possible identifications.

\section{Spectral fitting for \rx}
\label{sect:resultsrx}

For the model fits, the spectra of \hh\ were smoothed with
0.02\,\AA\ (FWHM) Gaussians to increase the signal-to-noise ratio
(S/N). It is then about 30 but deteriorates for $\lambda >
1500$\,\AA. In comparison to \hh, the UV flux of \rx\ is significantly
lower (about a factor of 150), but the exposure time of the G130M
spectrum was nine times higher, so the S/N is about four times
lower. To compensate for this, the spectrum was smoothed with a
0.05\,\AA\ Gaussian. The spectra of both stars look rather similar but
as a result of the lower S/N, fewer details can be seen in \rx. 

\subsection{Helium}

The lack of helium lines in the optical spectrum of \rx\ was claimed
as evidence for helium-deficiency. In fact, this conclusion is
difficult to draw. It has been shown that the most direct signature
would be the lack of a NLTE central emission reversal from
\ion{He}{ii} 4686\,\AA\ \citep{1991A&A...251..147W}. All other
\ion{He}{ii} lines are in absorption and are blended by \ion{C}{iv}
and \ion{O}{vi} lines at virtually equal wavelengths because all of these
ions are one-electron systems. The same holds for \ion{He}{ii}
1640\,\AA. In order to derive a strict upper limit for the He
abundance, a much better optical spectrum would be needed. 
\subsection{Carbon and oxygen; effective temperature}

Lines from \ion{C}{iv} and \ion{O}{vi} are identified and their
relative strengths are similar to those in \hh, such that we conclude
C/O =~1. However, there is a significant difference. The
\ion{O}{v}~1371\,\AA\ line is lacking, indicating a higher temperature
compared to \hh. A model with \Teff = 220\,000\,K (at \logg = 8) is
sufficiently hot to make the line so weak that it is not detectable in
the observation. On the other hand, an upper limit to \Teff\ can be
derived by the fact that the \ion{C}{iv} and \ion{O}{vi} lines become
weaker with increasing temperature. We have computed models with
250\,000\,K and 300\,000\,K. The hottest model is excluded because the
lines are too weak. We finally adopt \Teff = $250\,000\pm30\,000$\,K. 

In the models with \Teff = 200\,000, the \ion{C}{iv}
1548/1551\,\AA\ resonance doublet is comparable in strength with the
higher excited subordinate lines, as it is observed in \hh. At higher
temperatures, the resonance doublet becomes even \emph{\textup{weaker}} than
the subordinate lines and it turns into emission at \Teff $>$
250\,000\,K. This explains the non-detection of this line in the
HST/STIS spectrum taken by \cite{1999A&A...351L..27V}, albeit one has
to stress that the S/N ratio of that spectrum is very poor anyhow. 

\subsection{Nitrogen}

The \ion{N}{v} 1239/1243\,\AA\ resonance doublet is not
detectable. This is in stark contrast to the detection of two
\ion{N}{v} emission lines in the optical spectrum claimed by
\citet[][their Fig.\,1]{2002ASPC..261..653R}. Indeed, one of these features near
4340\,\AA\ is a \ion{Ne}{viii} line that is usually visible in the
hottest (\Teff $>$ 150\,000\,K) PG\,1159 stars
\citep{2007A&A...474..591W} and predicted by our \Teff = 250\,000
model for \rx. For the other feature at 4945\,\AA, however, we cannot
offer an alternative identification. From the absence of the
\ion{N}{v} UV resonance doublet, we determine an upper abundance limit
of N $<$ 0.001 for all models in the range \Teff =
200\,000--300\,000\,K. We found that models with such a low abundance
do not exhibit detectable optical \ion{N}{v} emission lines. 

\subsection{Neon}

The wide \ion{Ne}{viii} absorption trough in \hh\ is just barely
visible in \rx, immediately pointing at a lower neon abundance. This
is not a temperature effect because the \ion{Ne}{viii} lines become
stronger with increasing \Teff. The neon abundance we deduce is Ne =
0.01, with an estimated uncertainty of a factor of three.

\begin{table}
\begin{center}
\caption{Results of spectral analyses of \hh\ and \rx.\tablefootmark{a} }
\label{tab:results} 
\small
\begin{tabular}{cccc}
\hline 
\hline 
\noalign{\smallskip}
                   &  \hh                & RX\,J0439.8          & Reference\tablefootmark{b} \\
\hline 
\noalign{\smallskip}
\Teff / kK         & $200\pm20$          & $250\pm30$           & (1), this paper\\
\logg / cm s$^{-2}$ & $8.0\pm0.5$         & $8.0\pm0.5$          & (2)   \\
\noalign{\smallskip}
$M / M_\odot$       & $0.83^{+0.19}_{-0.15}$ & $0.86^{+0.16}_{-0.13}$  &  this paper  \\
\noalign{\smallskip}
$d$ / kpc            & $0.67^{+0.29}_{-0.52}$ & $9.2^{+4.1}_{-7.2}$     &  this paper  \\
\noalign{\smallskip}
$|z|$ / kpc          & $0.48^{+0.21}_{-0.38}$ & $5.6^{+2.5}_{-4.3}$     &  this paper  \\
\noalign{\smallskip}
He                 & $<0.01$             &                      & (1)   \\
C                  & 0.46                & 0.50                 & (2), this paper   \\
N                  & $<3.0\times10^{-5}$  & $<1.0\times10^{-3}$   & this paper   \\
O                  & 0.46                & 0.50                 & (2), this paper   \\
Ne                 & 0.06                & 0.01                 & (3,4,5), this paper   \\
Na                 & $<0.1$              &                      & (4)    \\
Mg                 & 0.02                &                      & (4)    \\
Al                 & $<0.1$              &                      & (4)    \\
Si                 & $<2.6 \times 10^{-3}$&                      & this paper   \\
Ca                 & $<6.4 \times 10^{-5}$&                      & (6), this paper   \\
Fe                 & 0.0013              &                      & (7)   \\
\noalign{\smallskip} \hline
\end{tabular} 
\tablefoot{
\tablefoottext{a}{Abundances in mass fractions. Stellar
    mass $M$ and distance $d$ were derived from comparison with evolutionary
    tracks. The paramter $z$ is the distance from the Galactic disk.}
\tablefoottext{b}{The last column gives references to results for \hh;
    results for \rx\ from this paper.
(1) \cite{2004A&A...427..685W};
(2) \cite{1991A&A...251..147W};
(3) \cite{1999A&A...347L...9W};
(4) \cite{2004A&A...427..685W};
(5) \cite{2007A&A...474..591W};
(6) \cite{2008A&A...492L..43W};
(7) \cite{2010ApJ...719L..32W}.
}
} 
\end{center}
\end{table}

\begin{figure}[t]
   \resizebox{\hsize}{!}{\includegraphics{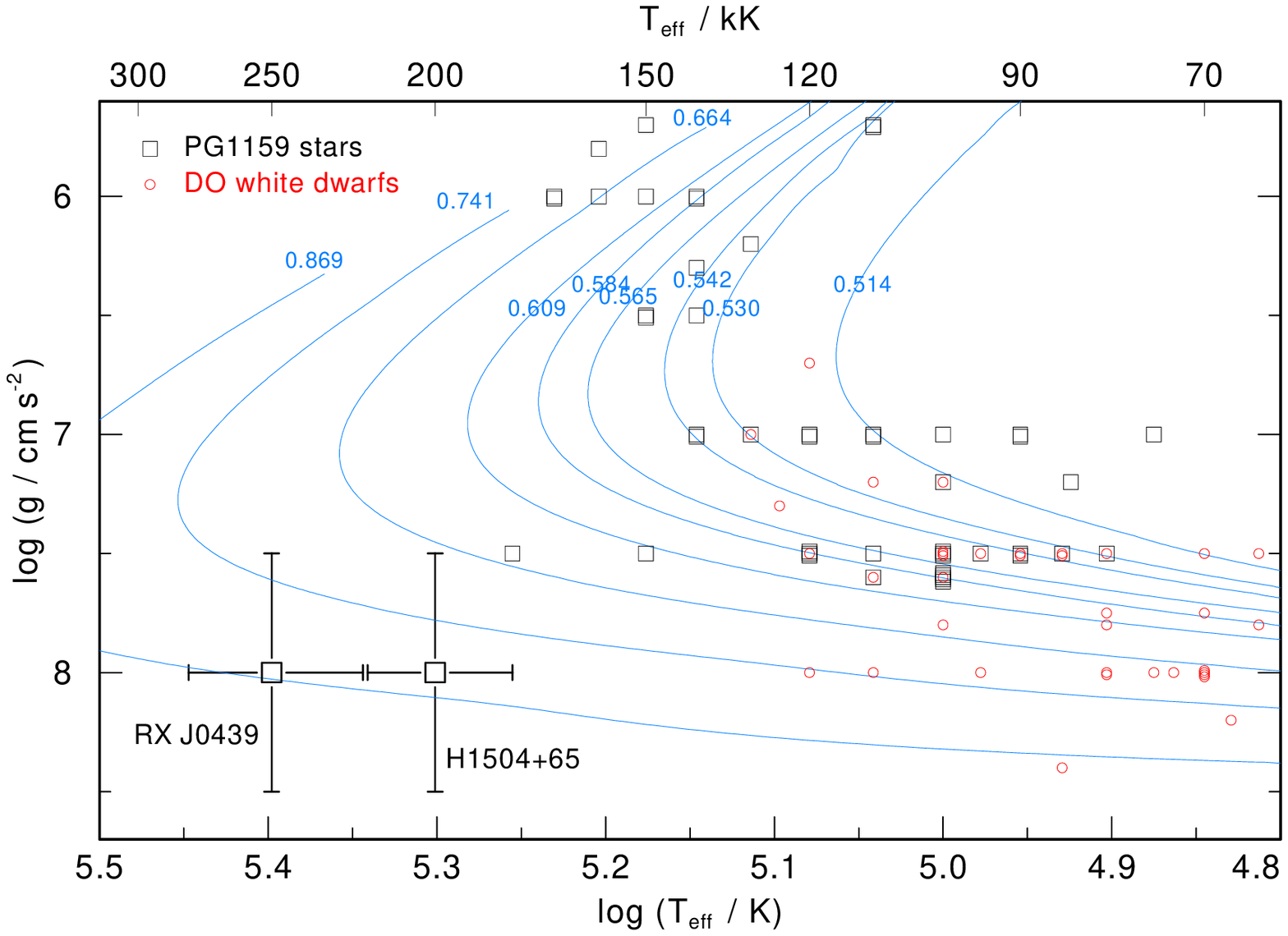}}
    \caption{Positions of \hh\ and \rx\ in the g--\Teff\ diagram.
      Also shown are the locations of PG\,1159 stars (squares) and DO
      white dwarfs (circles). The apparent clustering of stars at
        $\Delta$\logg = 0.5 steps is the consequence of employed model
        grids and analysis techniques that have typical errors of
        0.5\,dex \citep[see, e.g.][]{2006PASP..118..183W}.
        Evolutionary tracks by \citet{2009ApJ...704.1605A} are labelled
        with the stellar mass in solar units.  }
   \label{fig:gteff}
\end{figure}

\section{Stellar parameters and distances}
\label{sect:distances}

Table\,\ref{tab:results} summarises the results of our spectroscopic
analysis with results from previous work. The location of
both stars in the $g$--\Teff diagram is shown in
Fig.\,\ref{fig:gteff}. Also shown are evolutionary tracks for
hydrogen-deficient (pre-)white dwarfs computed by
\citet{2009ApJ...704.1605A} from which we derive the stellar
masses. The upper limits of the error ranges required extrapolation
from the two most massive tracks. We find mass ranges of
$0.68-1.02\,M_\odot$ and $0.73-1.02\,M_\odot$ for \hh\ and \rx,
respectively. The marginal difference between the mass value of
\hh\ derived here and in previous work mentioned in the Introduction
\citep{2004A&A...421.1169W} arises from improved evolutionary tracks.

It should be noted that the masses could suffer from
systematic errors. The used evolutionary tracks represent
hydrogen-deficient, He-shell burning post-AGB stars and their hot WD
descendants, whereas our stars in the present study are He-deficient.
Whatever the origin of the stars, the appropriate tracks would 
probably be different, although it may be hoped that the differences are
small at the advanced, WD evolutionary state.

\begin{table}
\begin{center}
\caption{Quantities used for distance determinations. }
\label{tab:magnitudes} 
\tiny
\begin{tabular}{ccccc}
\hline 
\hline 
\noalign{\smallskip}
Star & $E(B-V)$ & $V$ & $V_0$ & $H_\nu$\tablefootmark{e}\\
\hline 
\noalign{\smallskip}
\hh& 0.00\tablefootmark{a} & 16.24\tablefootmark{c} & 16.24 & $3.42\times10^{-3} $\\   
\rx& 0.09\tablefootmark{b} & 21.74\tablefootmark{d} & 21.65 & $4.30\times10^{-3} $\\     
\noalign{\smallskip} \hline
\end{tabular} 
\tablefoot{
\tablefoottext{a}{\cite{1998ApJ...502..858K};}
\tablefoottext{b}{derived from N$_{\rm H\,I}$, see text;}
\tablefoottext{c}{\cite{1986ApJ...309..230N};}
\tablefoottext{d}{\cite{1996A&A...307L..49V};}
\tablefoottext{e}{Eddington flux in erg cm$^{-2}$ s$^{-1}$ cm$^{-1}$. }
} 
\end{center}
\end{table}

From $g \sim M/R^2$ and $L \sim R^2 T_{\rm eff}^4$, we can determine
the stellar radius $R$ and luminosity $L$. We find that the radius of
both stars is $R\approx0.015\pm0.01\,R_\odot$. They are \emph{\textup{beyond}} the maximum
\Teff\ of their pre-WD evolution and their luminosity has dropped from
about $\log(L/L_\odot) = 4.5$ on the horizontal part of the post-AGB
evolutionary track in the HRD \citep[$M=0.869\,M_\odot$ track
  of][]{2009ApJ...704.1605A} to $\log(L/L_\odot) = 2.5$ and 2.9
(\hh\ and \rx, respectively).

Distances $d$ are computed by comparing the observed visual magnitudes
with model fluxes \citep[e.g.][]{1984A&A...130..119H}, as follows:
$$d {\rm (kpc)} = 7.1 \times \sqrt{H_\nu \cdot M \cdot 10^{0.4 V_0 -
    \log g} },$$ where $H_\nu$ is the model Eddington flux at
5400\,\AA, $M$ is the stellar mass in solar units, and $V_0$ is the
dereddened $V$ magnitude, $V_0 = V-3.2\times E(B-V)$. Reddening for
\rx\ is computed from the relation $\log ({\rm N}_{\rm H\,I}/E(B-V)) =
21.58$ \citep{1989A&AS...79..359G}, leading to $E(B-V)) = 0.09$, in
reasonable agreement with the value 0.06 derived from dust emission
maps\footnote{\url{http://irsa.ipac.caltech.edu/applications/DUST/}}
\citep{1998ApJ...500..525S}. The relevant quantities for the distance
determination are listed in Table\,\ref{tab:magnitudes}, and stellar
masses, surface gravities, and  the resulting distances are
listed in Table~\ref{tab:results}. 

The spectroscopic distance of \rx\ ($9.2^{+4.1}_{-7.2}$\,kpc)
clearly indicates that the star is a Galactic halo object and not
located in the LMC \citep[50\,kpc,][]{2013Natur.495...76P} as
suspected from its location, which is roughly 1 degree off the rim of
gaseous optical emission of the LMC
\citep{1994A&A...281L..61G}. Obviously, the direction is a chance
coincidence, just like the radial velocity of the star
($+220\pm10$\,km\,s$^{-1}$), which is close to that of the LMC \citep[+262
  km\,s$^{-1}$,][]{2012AJ....144....4M}. As mentioned in
Sect.\,\ref{sect:obs}, the interstellar lines in the \rx\ COS spectrum
 have a weak, high-velocity component ($+150\pm 10$\,km\,s$^{-1}$),
which can be attributed to a Galactic high-velocity cloud. There is no
component at the LMC systemic velocity.

\section{Summary and conclusions}
\label{sect:conclusions}      

\hh\ and \rx\ are the hottest known white dwarfs and they have carbon
and oxygen dominated atmospheres. Their extreme surface composition is
unique and the origin of this phenomenon remains mysterious. It may be
significant that their masses are considerably above the WD average
\citep[$0.696\pm0.010\,M_\odot$ for helium-atmosphere; i.e. DB white
  dwarfs,][]{2015MNRAS.446.4078K}. Our analyses yields a maximum mass
of $1.02\,M_\odot$ for both stars, which is rather close to the
minimum mass for ONeMg white dwarfs that descended from carbon-burning
super-AGB stars
\citep[$1.06\,M_\odot$;][]{2015MNRAS.446.2599D}. Hence, they could be
single high-mass white dwarfs, which for an unknown reason were eroded
down to their C/O envelope. Likewise, both stars could be usual CO
white dwarfs that underwent a late He-shell flash consuming the H-rich
envelope and, again for an unknown reason \citep[e.g.  extreme mass
  loss;][]{1992LNP...401..305S,2009A&A...494.1021A}, stripped their
He-envelope to expose their CO core. In both scenarios, the mass of
the main-sequence progenitor is relatively large, which is difficult
to reconcile with the position of \rx\ in the halo, unless it was
ejected from the disk (e.g. by the SN explosion of a binary
companion). Alternatively, the masses of \hh\ and \rx\  are high
enough that they could be the result of the coalescence of two CO
white dwarfs (with minimum masses of $\approx 0.5\,M_\odot$) whose H-
and He-deficiency results from the merging process.

While \hh\ is a relatively nearby WD ($d = 0.67$\,kpc), \rx\ belongs
to the Galactic halo ($|z| = 5.6$\,kpc). Its distance is 9.2\,kpc and,
thus, it is not a member of the LMC as previously assumed. The lack of
ISM lines with LMC systemic velocity corroborates this result. Its
persistent soft X-ray luminosity as well as its optical
non-variability \citep{1999A&A...351L..27V} are consistent with its
nature as a single, hot white dwarf that has no active nuclear burning
and is on the hot end of its WD cooling track. Consequently, its
cooling rate, according to the $0.869\,M_\odot$ post-AGB evolutionary
track of \citet{2009ApJ...704.1605A}, is rather slow
(d$T\mathrm{\hspace*{-0.4ex}_{eff}}$/dt = 38 K yr$^{-1}$), and
therefore, a 1\% change in the UV/optical flux level takes 66 years,
preventing a contemporary detection. This also explains the constant
X-ray flux measured by ROSAT over three years instead of a steadily
increasing X-ray flux. That was regarded as a challenge to stellar
evolution theory \citep{1999A&A...351L..27V}, however, it was based on
the erroneous assumption that the star is still a very luminous, fast
evolving pre-WD \emph{\textup{before}} entering the WD cooling
sequence.

\begin{acknowledgements} 
T. Rauch was supported by the German Aerospace Center (DLR) under
grant 05\,OR\,1402. This research has made use of the SIMBAD database,
operated at CDS, Strasbourg, France, and of NASA's Astrophysics Data
System Bibliographic Services. Some of the data presented in this
paper were obtained from the Mikulski Archive for Space Telescopes
(MAST). 
\end{acknowledgements}

\bibliographystyle{aa}
\bibliography{aa}

\begin{thebibliography}{50}
\expandafter\ifx\csname natexlab\endcsname\relax\def\natexlab#1{#1}\fi

\bibitem[{{Althaus} {et~al.}(2009{\natexlab{a}}){Althaus}, {C{\'o}rsico},
  {Torres}, \& {Garc{\'{\i}}a-Berro}}]{2009A&A...494.1021A}
{Althaus}, L.~G., {C{\'o}rsico}, A.~H., {Torres}, S., \& {Garc{\'{\i}}a-Berro},
  E. 2009{\natexlab{a}}, \aap, 494, 1021

\bibitem[{{Althaus} {et~al.}(2009{\natexlab{b}}){Althaus}, {Panei}, {Miller
  Bertolami}, {Garc{\'{\i}}a-Berro}, {C{\'o}rsico}, {Romero}, {Kepler}, \&
  {Rohrmann}}]{2009ApJ...704.1605A}
{Althaus}, L.~G., {Panei}, J.~A., {Miller Bertolami}, M.~M., {et~al.}
  2009{\natexlab{b}}, \apj, 704, 1605

\bibitem[{{Anderson}(1989)}]{1989ApJ...339..558A}
{Anderson}, L.~S. 1989, \apj, 339, 558

\bibitem[{{Bashkin} \& {Stoner}(1975)}]{1975aelg.book.....B}
{Bashkin}, S. \& {Stoner}, J.~O. 1975, {Atomic energy levels and Grotrian
  Diagrams - Vol.1: Hydrogen I - Phosphorus XV; Vol.2: Sulfur I - Titanium
  XXII}

\bibitem[{{Dere} {et~al.}(1997){Dere}, {Landi}, {Mason}, {Monsignori Fossi}, \&
  {Young}}]{1997A&AS..125..149D}
{Dere}, K.~P., {Landi}, E., {Mason}, H.~E., {Monsignori Fossi}, B.~C., \&
  {Young}, P.~R. 1997, \aaps, 125, 149

\bibitem[{{Doherty} {et~al.}(2015){Doherty}, {Gil-Pons}, {Siess}, {Lattanzio},
  \& {Lau}}]{2015MNRAS.446.2599D}
{Doherty}, C.~L., {Gil-Pons}, P., {Siess}, L., {Lattanzio}, J.~C., \& {Lau},
  H.~H.~B. 2015, \mnras, 446, 2599

\bibitem[{{Dufour} {et~al.}(2011){Dufour}, {B{\'e}land}, {Fontaine}, {Chayer},
  \& {Bergeron}}]{2011ApJ...733L..19D}
{Dufour}, P., {B{\'e}land}, S., {Fontaine}, G., {Chayer}, P., \& {Bergeron}, P.
  2011, \apjl, 733, L19

\bibitem[{{Dufour} {et~al.}(2007){Dufour}, {Liebert}, {Fontaine}, \&
  {Behara}}]{2007Natur.450..522D}
{Dufour}, P., {Liebert}, J., {Fontaine}, G., \& {Behara}, N. 2007, \nat, 450,
  522

\bibitem[{{G{\"a}nsicke} {et~al.}(2010){G{\"a}nsicke}, {Koester}, {Girven},
  {Marsh}, \& {Steeghs}}]{2010Sci...327..188G}
{G{\"a}nsicke}, B.~T., {Koester}, D., {Girven}, J., {Marsh}, T.~R., \&
  {Steeghs}, D. 2010, Science, 327, 188

\bibitem[{{Garc{\'{\i}}a-Berro} {et~al.}(1997){Garc{\'{\i}}a-Berro}, {Ritossa},
  \& {Iben}}]{1997ApJ...485..765G}
{Garc{\'{\i}}a-Berro}, E., {Ritossa}, C., \& {Iben}, Jr., I. 1997, \apj, 485,
  765

\bibitem[{{Greiner} {et~al.}(1994){Greiner}, {Hasinger}, \&
  {Thomas}}]{1994A&A...281L..61G}
{Greiner}, J., {Hasinger}, G., \& {Thomas}, H.-C. 1994, \aap, 281, L61

\bibitem[{{Groenewegen} \& {Lamers}(1989)}]{1989A&AS...79..359G}
{Groenewegen}, M.~A.~T. \& {Lamers}, H.~J.~G.~L.~M. 1989, \aaps, 79, 359

\bibitem[{{Heber} {et~al.}(1984){Heber}, {Hunger}, {Jonas}, \&
  {Kudritzki}}]{1984A&A...130..119H}
{Heber}, U., {Hunger}, K., {Jonas}, G., \& {Kudritzki}, R.~P. 1984, \aap, 130,
  119

\bibitem[{{Hummer} {et~al.}(1993){Hummer}, {Berrington}, {Eissner}, {Pradhan},
  {Saraph}, \& {Tully}}]{1993A&A...279..298H}
{Hummer}, D.~G., {Berrington}, K.~A., {Eissner}, W., {et~al.} 1993, \aap, 279,
  298

\bibitem[{{Jahn} {et~al.}(2007){Jahn}, {Rauch}, {Reiff}, {Werner}, {Kruk}, \&
  {Herwig}}]{2007A&A...462..281J}
{Jahn}, D., {Rauch}, T., {Reiff}, E., {et~al.} 2007, \aap, 462, 281

\bibitem[{{Kepler} {et~al.}(2015){Kepler}, {Pelisoli}, {Koester}, {Ourique},
  {Kleinman}, {Romero}, {Nitta}, {Eisenstein}, {Costa}, {K{\"u}lebi}, {Jordan},
  {Dufour}, {Giommi}, \& {Rebassa-Mansergas}}]{2015MNRAS.446.4078K}
{Kepler}, S.~O., {Pelisoli}, I., {Koester}, D., {et~al.} 2015, \mnras, 446,
  4078

\bibitem[{{Koester} {et~al.}(1982){Koester}, {Weidemann}, \&
  {Zeidler}}]{1982A&A...116..147K}
{Koester}, D., {Weidemann}, V., \& {Zeidler}, E.-M. 1982, \aap, 116, 147

\bibitem[{{Kramida} {et~al.}(1999){Kramida}, {Bastin}, {Bi{\'e}mont}, {Dumont},
  \& {Garnir}}]{1999JOSAB..16.1966K}
{Kramida}, A.~E., {Bastin}, T., {Bi{\'e}mont}, E., {Dumont}, P.-D., \&
  {Garnir}, H.-P. 1999, Journal of the Optical Society of America B Optical
  Physics, 16, 1966

\bibitem[{{Kruk} \& {Werner}(1998)}]{1998ApJ...502..858K}
{Kruk}, J.~W. \& {Werner}, K. 1998, \apj, 502, 858

\bibitem[{{Kurucz}(1991)}]{kurucz1991}
{Kurucz}, R.~L. 1991, in NATO ASIC Proc. 341: Stellar Atmospheres - Beyond
  Classical Models, ed. L.~{Crivellari}, I.~{Hubeny}, \& D.~G. {Hummer}, 441

\bibitem[{{Kurucz}(2009)}]{kurucz2009}
{Kurucz}, R.~L. 2009, in American Institute of Physics Conference Series, Vol.
  1171, American Institute of Physics Conference Series, ed. I.~{Hubeny}, J.~M.
  {Stone}, K.~{MacGregor}, \& K.~{Werner}, 43

\bibitem[{{Kurucz}(2011)}]{kurucz2011}
{Kurucz}, R.~L. 2011, Canadian Journal of Physics, 89, 417

\bibitem[{{Landi} {et~al.}(2013){Landi}, {Young}, {Dere}, {Del Zanna}, \&
  {Mason}}]{2013ApJ...763...86L}
{Landi}, E., {Young}, P.~R., {Dere}, K.~P., {Del Zanna}, G., \& {Mason}, H.~E.
  2013, \apj, 763, 86

\bibitem[{{McConnachie}(2012)}]{2012AJ....144....4M}
{McConnachie}, A.~W. 2012, \aj, 144, 4

\bibitem[{{Nousek} {et~al.}(1986){Nousek}, {Shipman}, {Holberg}, {Liebert},
  {Pravdo}, {White}, \& {Giommi}}]{1986ApJ...309..230N}
{Nousek}, J.~A., {Shipman}, H.~L., {Holberg}, J.~B., {et~al.} 1986, \apj, 309,
  230

\bibitem[{{Nugent} {et~al.}(1983){Nugent}, {Jensen}, {Nousek}, {Garmire},
  {Mason}, {Walter}, {Bowyer}, {Stern}, \& {Riegler}}]{1983ApJS...51....1N}
{Nugent}, J.~J., {Jensen}, K.~A., {Nousek}, J.~A., {et~al.} 1983, \apjs, 51, 1

\bibitem[{{Pietrzy{\'n}ski} {et~al.}(2013){Pietrzy{\'n}ski}, {Graczyk},
  {Gieren}, {Thompson}, {Pilecki}, {Udalski}, {Soszy{\'n}ski}, {Koz{\l}owski},
  {Konorski}, {Suchomska}, {Bono}, {Moroni}, {Villanova}, {Nardetto},
  {Bresolin}, {Kudritzki}, {Storm}, {Gallenne}, {Smolec}, {Minniti}, {Kubiak},
  {Szyma{\'n}ski}, {Poleski}, {Wyrzykowski}, {Ulaczyk}, {Pietrukowicz},
  {G{\'o}rski}, \& {Karczmarek}}]{2013Natur.495...76P}
{Pietrzy{\'n}ski}, G., {Graczyk}, D., {Gieren}, W., {et~al.} 2013, \nat, 495,
  76

\bibitem[{{Rauch} \& {Deetjen}(2003)}]{2003ASPC..288..103R}
{Rauch}, T. \& {Deetjen}, J.~L. 2003, in Astronomical Society of the Pacific
  Conference Series, Vol. 288, Stellar Atmosphere Modeling, ed. I.~{Hubeny},
  D.~{Mihalas}, \& K.~{Werner}, 103

\bibitem[{{Reindl} {et~al.}(2014){Reindl}, {Rauch}, {Werner}, {Kruk}, \&
  {Todt}}]{2014A&A...566A.116R}
{Reindl}, N., {Rauch}, T., {Werner}, K., {Kruk}, J.~W., \& {Todt}, H. 2014,
  \aap, 566, A116

\bibitem[{{Reinsch} {et~al.}(2002){Reinsch}, {Beuermann}, \&
  {G{\"a}nsicke}}]{2002ASPC..261..653R}
{Reinsch}, K., {Beuermann}, K., \& {G{\"a}nsicke}, B.~T. 2002, in Astronomical
  Society of the Pacific Conference Series, Vol. 261, The Physics of
  Cataclysmic Variables and Related Objects, ed. B.~T. {G{\"a}nsicke},
  K.~{Beuermann}, \& K.~{Reinsch}, 653

\bibitem[{{Richter} {et~al.}(1999){Richter}, {de Boer}, {Widmann},
  {Kappelmann}, {Gringel}, {Grewing}, \& {Barnstedt}}]{1999Natur.402..386R}
{Richter}, P., {de Boer}, K.~S., {Widmann}, H., {et~al.} 1999, \nat, 402, 386

\bibitem[{{Schlegel} {et~al.}(1998){Schlegel}, {Finkbeiner}, \&
  {Davis}}]{1998ApJ...500..525S}
{Schlegel}, D.~J., {Finkbeiner}, D.~P., \& {Davis}, M. 1998, \apj, 500, 525

\bibitem[{{Sch{\"o}nberner} \& {Bl{\"o}cker}(1992)}]{1992LNP...401..305S}
{Sch{\"o}nberner}, D. \& {Bl{\"o}cker}, T. 1992, in Lecture Notes in Physics,
  Berlin Springer Verlag, Vol. 401, The Atmospheres of Early-Type Stars, ed.
  U.~{Heber} \& C.~S. {Jeffery}, 305

\bibitem[{{Seaton} {et~al.}(1994){Seaton}, {Yan}, {Mihalas}, \&
  {Pradhan}}]{1994MNRAS.266..805S}
{Seaton}, M.~J., {Yan}, Y., {Mihalas}, D., \& {Pradhan}, A.~K. 1994, \mnras,
  266, 805

\bibitem[{{Smoker} {et~al.}(2015){Smoker}, {Fox}, \&
  {Keenan}}]{2015MNRAS.451.4346S}
{Smoker}, J.~V., {Fox}, A.~J., \& {Keenan}, F.~P. 2015, \mnras, 451, 4346

\bibitem[{{van Teeseling} {et~al.}(1999){van Teeseling}, {G{\"a}nsicke},
  {Beuermann}, {Dreizler}, {Rauch}, \& {Reinsch}}]{1999A&A...351L..27V}
{van Teeseling}, A., {G{\"a}nsicke}, B.~T., {Beuermann}, K., {et~al.} 1999,
  \aap, 351, L27

\bibitem[{{van Teeseling} {et~al.}(1996){van Teeseling}, {Reinsch}, \&
  {Beuermann}}]{1996A&A...307L..49V}
{van Teeseling}, A., {Reinsch}, K., \& {Beuermann}, K. 1996, \aap, 307, L49

\bibitem[{{Werner}(1991)}]{1991A&A...251..147W}
{Werner}, K. 1991, \aap, 251, 147

\bibitem[{{Werner} {et~al.}(2003){Werner}, {Deetjen}, {Dreizler}, {Nagel},
  {Rauch}, \& {Schuh}}]{2003ASPC..288...31W}
{Werner}, K., {Deetjen}, J.~L., {Dreizler}, S., {et~al.} 2003, in Astronomical
  Society of the Pacific Conference Series, Vol. 288, Stellar Atmosphere
  Modeling, ed. I.~{Hubeny}, D.~{Mihalas}, \& K.~{Werner}, 31

\bibitem[{{Werner} \& {Dreizler}(1999)}]{1999JCoAM.109...65W}
{Werner}, K. \& {Dreizler}, S. 1999, Journal of Computational and Applied
  Mathematics, 109, 65

\bibitem[{{Werner} {et~al.}(2012){Werner}, {Dreizler}, \& {Rauch}}]{tmap2012}
{Werner}, K., {Dreizler}, S., \& {Rauch}, T. 2012, {TMAP: T{\"u}bingen NLTE
  Model-Atmosphere Package}, Astrophysics Source Code Library

\bibitem[{{Werner} \& {Herwig}(2006)}]{2006PASP..118..183W}
{Werner}, K. \& {Herwig}, F. 2006, \pasp, 118, 183

\bibitem[{{Werner} {et~al.}(2004{\natexlab{a}}){Werner}, {Rauch}, {Barstow}, \&
  {Kruk}}]{2004A&A...421.1169W}
{Werner}, K., {Rauch}, T., {Barstow}, M.~A., \& {Kruk}, J.~W.
  2004{\natexlab{a}}, \aap, 421, 1169

\bibitem[{{Werner} {et~al.}(2014){Werner}, {Rauch}, \&
  {Kepler}}]{2014A&A...564A..53W}
{Werner}, K., {Rauch}, T., \& {Kepler}, S.~O. 2014, \aap, 564, A53

\bibitem[{{Werner} {et~al.}(2007){Werner}, {Rauch}, \&
  {Kruk}}]{2007A&A...474..591W}
{Werner}, K., {Rauch}, T., \& {Kruk}, J.~W. 2007, \aap, 474, 591

\bibitem[{{Werner} {et~al.}(2008){Werner}, {Rauch}, \&
  {Kruk}}]{2008A&A...492L..43W}
{Werner}, K., {Rauch}, T., \& {Kruk}, J.~W. 2008, \aap, 492, L43

\bibitem[{{Werner} {et~al.}(2010){Werner}, {Rauch}, \&
  {Kruk}}]{2010ApJ...719L..32W}
{Werner}, K., {Rauch}, T., \& {Kruk}, J.~W. 2010, \apjl, 719, L32

\bibitem[{{Werner} {et~al.}(2004{\natexlab{b}}){Werner}, {Rauch}, {Reiff},
  {Kruk}, \& {Napiwotzki}}]{2004A&A...427..685W}
{Werner}, K., {Rauch}, T., {Reiff}, E., {Kruk}, J.~W., \& {Napiwotzki}, R.
  2004{\natexlab{b}}, \aap, 427, 685

\bibitem[{{Werner} \& {Wolff}(1999)}]{1999A&A...347L...9W}
{Werner}, K. \& {Wolff}, B. 1999, \aap, 347, L9

\bibitem[{{Woosley} \& {Heger}(2015)}]{2015ApJ...810...34W}
{Woosley}, S.~E. \& {Heger}, A. 2015, \apj, 810, 34

\end{thebibliography}

\end{document}